# Nanoscale structural and electrical properties of graphene grown on AlGaN by catalyst-free chemical vapor deposition

F. Giannazzo [1,*], R. Dagher [2], E. Schilirò [1], S. E. Panasci [1,3], G. Greco [1], G. Nicotra [1], F. Roccaforte [1], S. Agnello [4,1], J. Brault [2], Y. Cordier [2], A. Michon [2]

[1] *Consiglio Nazionale delle Ricerche – Istituto per la Microelettronica e Microsistemi (CNR-IMM), Strada VIII, n. 5 Zona Industriale, 95121 Catania, Italy*

[2] *Université Côte d'Azur, CNRS, CRHEA, Rue Bernard Grégory, 06560 Valbonne, France*

[3] *Department of Physics and Astronomy, University of Catania, via Santa Sofia 64, 95123 Catania, Italy*

[4] *Department of Physics and Chemistry "E. Segrè", University of Palermo, via Archirafi 36, 90123 Palermo, Italy*

\* *Correspondence: filippo.giannazzo@imm.cnr.it*

**Abstract:** The integration of graphene (Gr) with nitride semiconductors is highly interesting for applications in high-power/high-frequency electronics and optoelectronics. In this work, we demonstrated the direct growth of Gr on $Al_{0.5}Ga_{0.5}N$/sapphire templates by propane ($C_3H_8$) chemical vapor deposition (CVD) at temperature of 1350°C. After optimization of the $C_3H_8$ flow rate, a uniform and conformal Gr coverage was achieved, which proved beneficial to prevent degradation of AlGaN morphology. X-ray photoemission spectroscopy (XPS) revealed Ga loss and partial oxidation of Al in the near-surface AlGaN region. Such chemical modification of a ~2 nm thick AlGaN surface region was confirmed by cross-sectional scanning transmission electron microscopy (STEM) combined with electron energy loss spectroscopy (EELS), which also showed the presence of a bilayer of Gr with partial $sp^2/sp^3$ hybridization. Raman spectra indicated that the deposited Gr is nanocrystalline (with domain size ~7 nm) and compressively strained. A Gr sheet resistance of ~15.8 kΩ/sq was evaluated by four-point-probe measurements, consistently with the nanocrystalline nature of these films. Furthermore, nanoscale resolution current mapping by conductive atomic force microscopy (C-AFM) indicated local variations of the Gr carrier density at a mesoscopic scale, which can be ascribed to changes in the charge transfer from the substrate due to local oxidation of AlGaN or to the presence of Gr wrinkles.

## 1. Introduction

The integration of Gr and related two-dimensional (2D) materials with bulk semiconductors has been the object of intensive investigations in the last years. This approach presents the advantage of combining the functional properties of Gr with the well-assessed electronic quality of semiconductor substrates, and it currently represents a viable root towards the industrial exploitation of Gr in electronics/optoelectronics [1,2].

To date, several efforts have been done to integrate Gr with silicon, which still represents the dominant platform for digital and low-power electronics. On the other hand, group III-Nitride (III-N) semiconductors (including GaN, AlN, InN and their alloys) are widely employed in many optoelectronics and power electronic devices. The integration of Gr with these materials has been also explored by different research groups, with the aim to improve the performances of existing GaN-based devices, as well as to demonstrate novel device concepts [3,4,5]. As an example, due to its excellent electronic transport properties [6,7,8,9] and high optical transparency (≈97.7% from UV to near-IR) [10,11], Gr has been considered as a transparent conductive electrode for GaN light emitting diodes (LEDs) [12,13,14,15] in replacement to currently used indium-tin-oxide (ITO). Thanks to its excellent thermal conductivity (up to 5000 W m$^{-1}$K$^{-1}$) [16], Gr has been also proposed as a suitable candidate to address self-heating problems in high-power HEMT devices based on AlGaN/GaN heterostructures [17], as well as in high power solid state optoelectronic devices [18]. Finally, the ultimate single atomic thickness of a Gr electrode (allowing ballistic electronic transit in its transversal direction), combined with the excellent rectifying properties of the Gr junction with Al(Ga)N/GaN heterostructures, have been recently exploited to implement vertical hot electron transistors (HETs) for ultra-high-frequency (THz) applications [5,19]. Besides device fabrication, single or few layers of Gr have been also employed as compliant interlayers to reduce the dislocation density of GaN films grown by metal organic chemical vapor deposition (MOCVD) on sapphire [20] or SiC [21]. Furthermore, Gr interlayers have been also used to grow good quality GaN on (100) oriented Si substrates (commonly used in the fabrication of Si electronic devices), which is an important step towards monolithic integration of GaN with Si-based complementary-metal-oxide-semiconductor (CMOS) technology [22].

To date, the most used approach for Gr integration has been the chemical vapor deposition (CVD) of Gr on catalytic metals (typically Ni or Cu in the form of polycrystalline thin films or foils) [23,24], followed by its transfer to the semiconductor surface. The use of catalytic substrates in the Gr CVD growth presents the advantage of lowering the energy barrier for the dissociation of carbon precursors, allowing Gr formation at temperatures in the order of 1000°C. The post-growth transfer procedure typically requires the use of a protective polymeric layer, such as PMMA, onto Gr to enable handling of this ultrathin membrane. Detachment of Gr from the native substrate is obtained either by complete chemical etching of the metal or by Gr delamination using electrolytic methods [25]. The transfer of the polymer/Gr stack on the target substrate is carried out, either by fishing, printing or roll-to-roll [25,26]. Finally, removal of the polymeric carrier layer from Gr surface is performed by using proper solvents, eventually followed by thermal treatments to eliminate polymer residuals. Although transfer is a versatile and widely used method for Gr integration with arbitrary substrates, it can suffer from some drawbacks related to Gr damage (cracks, wrinkles, folding) during handling and from undesired contaminations, including metal contaminations [27] originating from the catalytic metal substrate, that are only partially reduced by Gr delamination without substrate etching [25]. Furthermore, the final device structure can suffer from a lack of robustness, due to adhesion problems between transferred Gr and the substrate. In the last years, progresses on transfer approaches of Gr onto GaN or $Al_xGa_{1-x}N$/GaN heterostructures have been reported [5,28].

As an alternative approach to Gr transfer, the direct deposition of Gr on the III-N substrates would be highly desirable. However, CVD growth of Gr from carbon precursors on these non-catalytic substrates represents a challenging task, as it requires significantly higher temperatures as compared to conventional deposition on metal catalysts. The need of high deposition temperatures represents a severe limitation for Gr CVD on GaN, which is known to undergo decomposition and strong degradation of surface morphology at temperatures in the order of 1000 °C [29]. For this reason, only few reports can be found in the literature on the direct Gr growth on GaN. As an example, Sun et al. [30] developed a CVD process for the deposition of Gr on GaN templates at a temperature of 950 °C with $C_2H_2$/$NH_3$ gases, where $NH_3$ played the dual role of compensating the loss of nitrogen from GaN and to release $H_2$, which is helpful for Gr growth. The deposited carbon films (with 2-4 nm thickness) were uniform, transparent and conducting, but their sheet resistance was significantly higher as compared to standard Gr. More recently, Kim et al. [31] employed plasma-enhanced CVD (PECVD) with $CH_4$/$H_2$ at a temperature of 600 °C to directly grow polycrystalline Gr films on the p-GaN topmost layer of an LED structure. Although the PECVD Gr on GaN showed significant structural disorder (as indicated by Raman spectra),

LED devices with these directly grown Gr electrodes exhibited very encouraging electrical properties as compared to the LEDs fabricated with transferred Gr electrodes [31].

Differently than GaN, thin films and templates of AlN are typically able to sustain high thermal budgets without degradation [32]. The first experiments on Gr CVD growth onto AlN were carried out using AlN templates on a Si (111) substrate and propane ($C_3H_8$) as the carbon source [33]. More recently, the possibility of depositing few layers of Gr on bulk AlN (Al and N face) and on AlN templates grown on SiC, without significantly degrading the morphology of AlN substrates/templates, has been also demonstrated [34]. The direct CVD growth of Gr on an ultra-wide-bandgap semiconductor like AlN can open interesting perspectives in optoelectronics, e.g. in deep-UV LEDs technology. In this context, it would be very interesting to explore the possibility of Gr deposition also on $Al_xGa_{1-x}N$ alloys with high Al content (x>0.5). However, to the best of our knowledge, studies on the morphological/chemical stability of these Al-rich AlGaN surfaces under the high temperature conditions used for non-catalytic CVD growth of Gr are missing.

In this paper, we report a detailed morphological, structural and electrical investigation of few layers Gr directly grown onto $Al_xGa_{1-x}N$ on sapphire templates (with x=0.5 and 0.65) by non-catalytic CVD. After optimization of the $C_3H_8$ flow rate, a uniform and conformal Gr coverage was achieved, which proved to be beneficial to prevent the morphological degradation of the underlying AlGaN surface. However, the loss of Ga and partial Al oxidation in the near-surface AlGaN region was observed by XPS and cross-sectional STEM/EELS analyses. These high resolution microscopic/spectroscopic measurements also showed the presence of a bilayer of Gr with partial $sp^2/sp^3$ hybridization. Raman spectra indicated that the deposited Gr bilayer is nanocrystalline and compressively strained. Furthermore, the electrical properties of this Gr membrane have been evaluated on macroscopic scale by four-point-probe measurement, showing a sheet resistance of ~15.8 kΩ/sq, and at nanoscale by conductive atomic force microscopy (C-AFM) analyses, showing local variations of the Gr carrier density, probably due to local changes in the charge transfer from the AlGaN template.

## 2. Materials and Methods

The AlGaN templates on sapphire used as substrates for Gr CVD deposition studies were grown in a molecular beam epitaxy (MBE) reactor. First a GaN buffer layer (BL) was grown on a sapphire wafer, followed by an AlN layer (100 nm to 200 nm thick) and, finally, by the topmost AlGaN layer with a thickness of 500 nm following the growth conditions described in [35,36,37]. Two $Al_xGa_{1-x}N$ templates with Al molar fractions x=0.5 and 0.65 have been prepared for our experiments, as schematically illustrated in Fig.1(a) and (e).

The as-grown samples were subsequently cleaved into 5×5 mm$^2$ pieces, used to perform preliminary thermal annealing trials and the Gr deposition experiments within a CVD reactor. The preliminary annealing experiments, carried out in N$_2$ ambient (10 slm of N$_2$ at 800 mbar for 5 minutes) and at temperatures ranging from 1250 °C to 1450 °C, were aimed to evaluate the morphological degradation of AlGaN surface due to thermal decomposition effects. Following this study, the temperature of 1350 °C was chosen as a suitable value to perform the CVD Gr growth. Therefore, the Gr deposition was carried out at this temperature for 5 minutes under 10 slm of N$_2$ at 800 mbar and different propane (C$_3$H$_8$) flow rates from 2 to 5 sccm.

The surface morphology of the AlGaN templates (before and after the thermal annealing and Gr growth) was systematically characterized by tapping mode atomic force microscopy (AFM) measurements performed using a DI3100 microscope with Nanoscope IV controller. Furthermore, the deposition of sp$^2$ hybridized carbon and the compositional changes of the near surface region of the AlGaN templates following the CVD Gr growth processes were monitored by X-ray photoelectron spectroscopy (XPS) analyses, performed using a Thermo Scientific K α system with Al$_{K\alpha}$ monochromated source.

Raman spectroscopy analyses of CVD-grown Gr were carried out using a Bruker SENTERRA spectrometer equipped with a confocal microscopy system and a 532 nm (2.33 eV) excitation laser at power lower than 5 mW and best spectral resolution 9 cm$^{-1}$.

High resolution structural and chemical characterization of Gr/AlGaN samples cross-sectioned by focused ion beam (FIB) was carried out using scanning transmission electron microscopy (STEM) and electron energy loss spectroscopy (EELS). These analyses were carried out with a sub-Angstrom aberration-corrected JEOL ARM200F atomic resolution microscope equipped with EELS spectrometer. Measurements were performed at the so-called gentle STEM condition [38] at 60-keV primary beam energy, which is lower than the knock-on threshold for carbon atoms (~85 keV), thus ensuring no damage on Gr during STEM and EELS acquisition.

After these morphological, structural and chemical investigations, the electrical properties of CVD-grown Gr onto AlGaN have been investigated both at macroscopic scale by four point probes sheet resistance measurements in the van der Pauw configuration, and at nanoscale by conductive atomic force microscopy (CAFM) current mapping using Pt-coated Si tips. The CAFM analyses were carried out with a DI3100 microscope with Nanoscope V controller.

## 3. Results and discussion

As discussed in the experimental section, some of the AlGaN samples were subjected to preliminary thermal treatments under $N_2$ flux in order to monitor thermal decomposition effects of the AlGaN surface and to identify the suitable temperature conditions for subsequent Gr CVD growth under $N_2/C_3H_8$ flux. Fig.1(b)-(d) show three typical AFM morphologies for the pristine $Al_{0.5}Ga_{0.5}N$ surface (b), and after annealing at 1250°C (c) and 1450 °C (d) under 10 slm of $N_2$ at 800 mbar for 5 minutes. Three AFM analyses performed with the same scan size on the pristine $Al_{0.65}Ga_{0.35}N$ sample and after identical thermal treatments are reported in Fig.1(f)-(h).

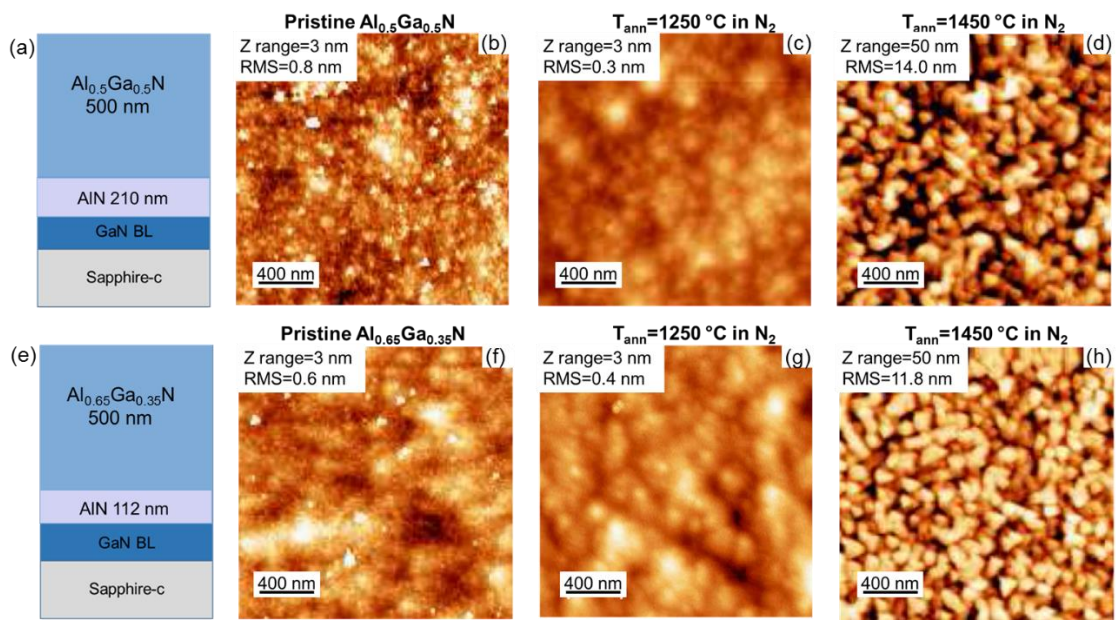

**Figure 1.** (a) Schematic cross section of the MBE grown $Al_{0.5}Ga_{0.5}N$ template on sapphire used for graphene CVD deposition. AFM images of the pristine AlGaN surface (b) and after thermal annealing under $N_2$ flux at 1250°C (c) and 1450°C (d). (e) Schematic cross section of the MBE grown $Al_{0.65}Ga_{0.65}N$ template on sapphire used for graphene CVD deposition. AFM images of the pristine AlGaN surface (f) and after thermal annealing under $N_2$ flux at 1250°C (g) and 1450°C (h).

The pristine $Al_xGa_{1-x}N$ surfaces exhibit root mean square roughness (RMS) values of 0.8 nm and 0.6 nm, respectively, for the templates with x=0.5 and 0.65 Al mole fraction. Such roughness values are due, in part, to the presence of aggregates on the AlGaN surface, originating from the lack of mobility of Al adatoms during MBE growth performed at 850-870°C [39]. A smoother morphology, with reduced RMS values of 0.3 nm and 0.4 nm, respectively, can be observed after annealing of the two samples at 1250°C, as shown in Fig.1 (c) and (g). On the other hand, after annealing at 1450°C under the same $N_2$ flux, the two AlGaN surfaces undergo a severe degradation with the appearance of a large density of pits, resulting in an increase of RMS up to 14 nm and 11.8 nm, respectively (as shown in Fig.1(d) and (h)).

These preliminary experiments revealed a similar evolution of the morphology with annealing temperature, independently of the Al content in the two AlGaN layers. In particular, the temperature of 1450 °C represents an upper limit for which strong decomposition of AlGaN surface occurs, whereas minor morphological modifications are observed at a temperature of 1250 °C. Based on these results, the $N_2/C_3H_8$ CVD growth experiments will be carried out at the intermediate temperature of 1350 °C, properly chosen to reduce AlGaN thermal decomposition and to provide, at the same time, the energy needed for Gr formation on this non-catalytic surface. It is worth noting that the same temperature has been recently employed to achieve optimal Gr growth onto AlN/SiC templates by a $N_2/C_3H_8$ CVD process [34].

In the following, the results of the morphological, structural and chemical analyses of the $Al_{0.5}Ga_{0.5}N$ template after the CVD process are reported.

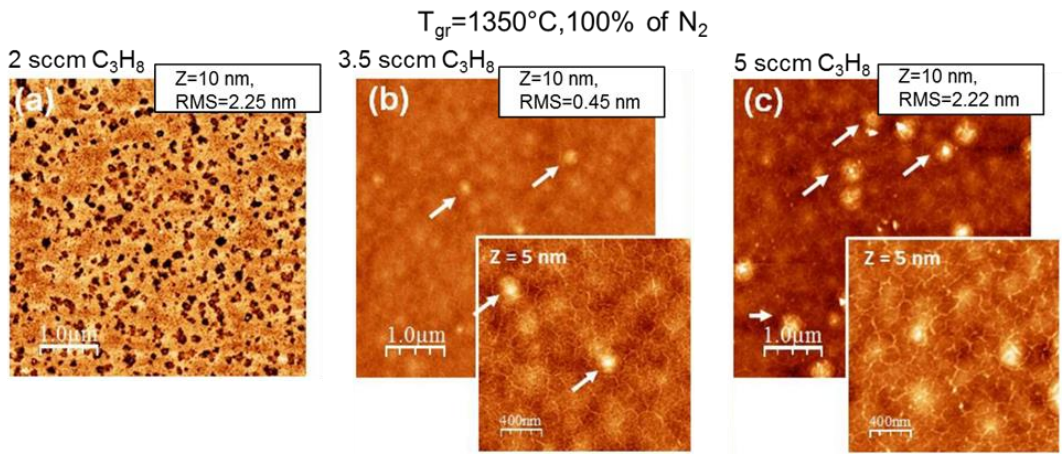

**Figure 2.** AFM images of the $Al_{0.5}Ga_{0.5}N$ surface after CVD depositions performed at 1350 °C for 5 minutes with 10 slm of $N_2$ at 800 mbar and different $C_3H_8$ flow rates of 2 sccm (a), 3.5 sccm (b) and 5 sccm (c). The higher magnification images in the inserts of (b) and (c) show the presence of a network of wrinkles, associated to the deposition of a carbon membrane, along with regions showing bright contrast (indicated by white arrows) associated to carbon over-deposition.

Figure 2 shows three typical AlGaN surface morphologies after the CVD process at 1350 °C for 5 minutes with 10 slm of $N_2$ at 800 mbar and $C_3H_8$ flow rates of 2 sccm (a), 3.5 sccm (b) and 5 sccm (c). A slight degradation of the AlGaN surface, with the appearance of small pits, can be observed after the CVD process with the 2 sccm $C_3H_8$ flow rate. Interestingly, by increasing the $C_3H_8$ flow rate to 3.5 sccm (Fig.2(b)), no pits associated to AlGaN decomposition/etching are observed. In addition to the improved morphology, the presence of a network of wrinkles on the AlGaN surface is evident in the higher magnification insert of Fig.2(b), along with regions showing bright contrast indicated by white arrows. The appearance of wrinkles can be an indication of the deposition of a carbon membrane, and the bright regions suggest overdeposition of carbon in some areas. We can speculate that the carbon membrane deposition acts as a capping layer, preventing AlGaN decomposition. By further increasing $C_3H_8$ flow rate to 5 sscm (Fig.2(c) and its insert), the wrinkles of the deposited carbon membrane appear to be more pronounced and the density of clusters associated to over-deposition of carbon increases.

The deposition of graphitic carbon was monitored using XPS, by comparing the C1s core-level spectra measured on the pristine AlGaN surface and after the CVD processes with different $C_3H_8$ flow rates, as shown in Figure 3. A low intensity peak at 284.7 eV, associated to adventitious carbon, can be observed for the as-grown AlGaN surface. After CVD with 2 sccm of $C_3H_8$, the C1s peak was downshifted to 284.4 eV, i.e. closer to the expected binding energy value for $sp^2$ carbon, suggesting the beginning of a graphitic phase deposition. Increasing the $C_3H_8$ flow rate to 3.5 sccm resulted in the increase of the intensity of the C1s peaks, located at ~284.5 eV.

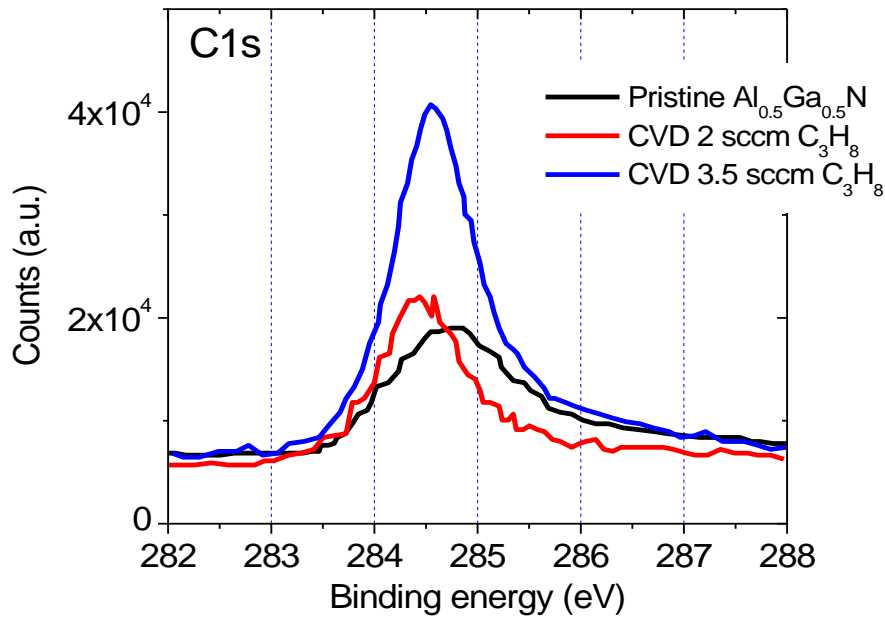

**Figure 3.** C1s core-level XPS spectra collected on the pristine $Al_{0.5}Ga_{0.5}N$ surface and after CVD depositions performed at 1350 °C for 5 minutes with 10 slm of $N_2$ at 800 mbar and different $C_3H_8$ flow rates of 2 sccm and 3.5 sccm.

Based on the combined results of the morphological analysis and XPS investigation in Figures 2 and 3, the CVD process with 3.5 sccm of $C_3H_8$ was selected as the optimal condition to achieve the formation of a graphitic carbon membrane, while avoiding formation of particles arising from carbon-overdeposition.

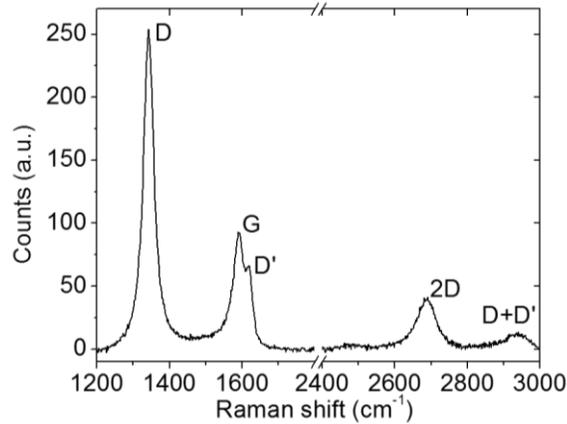

**Figure 4.** Typical Raman spectrum of the as-grown graphene on AlGaN at 1350 °C using optimized conditions (10 slm of $N_2$ at 800 mbar and a $C_3H_8$ flow rate of 3.5 sccm).

Figure 4 shows a typical Raman spectrum collected on the as-grown sample under these optimal process conditions. This spectrum features the characteristic peaks of graphitic carbon, i.e. the G and 2D peaks located at ~1593 and 2686 cm$^{-1}$, respectively. The 2D band exhibits a single Lorenzian shape with a large full-width-at-half-maximum (FWHM) of ~63.7 cm$^{-1}$, indicating the presence of few-layers of Gr with rotational disorder. Furthermore, the defect-related D, D', and D+D' peaks can be observed in this spectrum. In particular, the D' and D+D' peaks and the high D/G intensity ratio can be associated to a high density of grain boundaries between Gr domains [33]. Furthermore, an average lateral domain size L~7 nm was evaluated according to the relation $L=(2.4\times10^{-10})\lambda^4(I_D/I_G)^{-1}$, where $\lambda$= 532 nm is the wavelength of the laser probe and $I_D/I_G \approx 2.7$ is the intensity ratio of the D and G peaks [40]. Finally, the blue shift of both G and 2D peaks with respect to the positions for the ideal free-standing Gr (at 1582 cm$^{-1}$ and 2670 cm$^{-1}$) [41] can be attributed to the compressive strain of the deposited Gr membrane, confirmed also by the observation of wrinkles in the AFM images in Fig.2(b) and (c). Similarly to the case of CVD-grown Gr on other substrates, it can be supposed that the strain accumulates during the cooling down stage of the CVD process [42], due to the differences in thermal expansion coefficients between the graphitic membrane and the substrate.

The analyses reported so far indicate the formation of a nano-crystalline Gr membrane without significant morphological degradation of the underlying $Al_{0.5}Ga_{0.5}N$ surface under the optimized deposition conditions at 1350 °C with 3.5 sccm of $C_3H_8$ under $N_2$ flux.

Near surface chemical modifications of the AlGaN template during the CVD process have been investigated by comparing the XPS core-level peaks of Ga3d, Al2p, N1s and O1s acquired on the pristine AlGaN sample (reference) and after the Gr deposition, as shown in Figure 5(a)-(d).

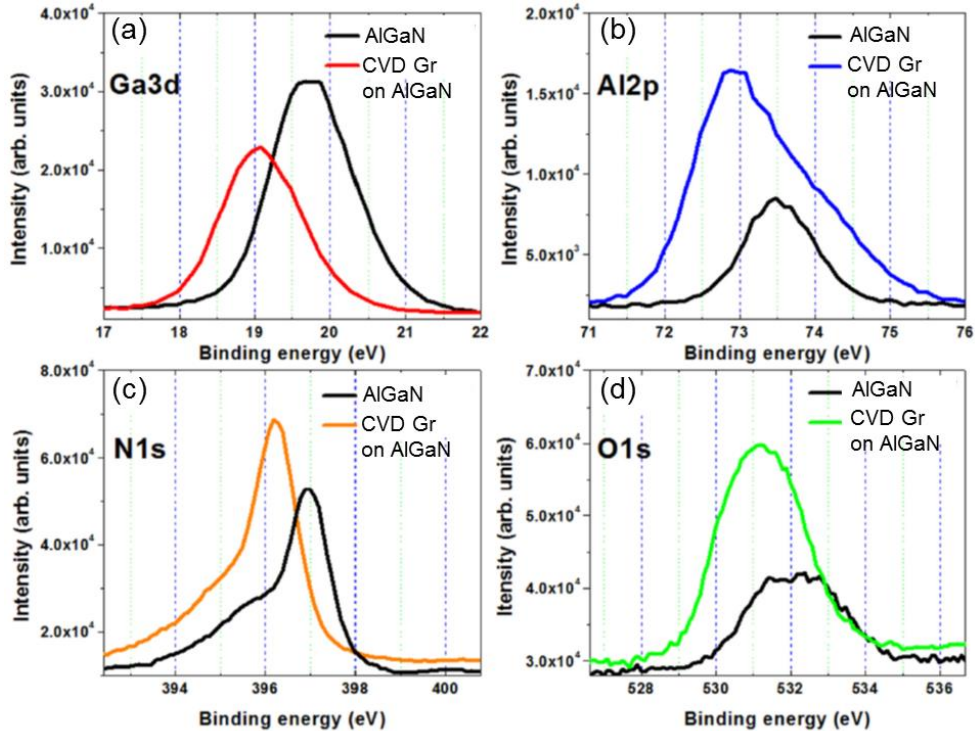

**Figure 5.** (a) Ga3d, (b) Al2p, (c) N1s and (d) O1s core-level peaks for the pristine AlGaN sample and after CVD with 3.5 sccm $C_3H_8$ at 1350°C under 10 slm of $N_2$ at 800 mbar.

After the CVD process, all the peaks exhibit a shift toward lower binding energies, as compared to the reference peaks for pristine AlGaN. This can be ascribed to an overall shift of the Fermi level of the system, after the deposition of the Gr membrane. Furthermore, we can observe a decrease in the intensity of the Ga3d peak (Fig.5(a)) accompanied by a relative increase in the intensity of the other peaks (Fig.5(b)-(d)). This behavior can be attributed to the preferential sublimation of Ga from the surface region during the CVD process at 1350 °C, resulting in an excess of Al. In addition, the Al2p peak after the Gr CVD growth exhibits an asymmetric shape towards higher binding energy, as compared to the symmetric peak measured on the reference AlGaN sample (see Fig.5(b)). This asymmetric shape can be ascribed to a new component due to the formation of Al bonds with oxygen, as the expected binding energy of $Al_2O_3$ is around 74.5 eV [43,44]. Partial oxidation of Al is also confirmed by the relative increase in the intensity of the O1s peak with respect to that of the native oxide present on the pristine

AlGaN surface (Fig.5(d)). The formation of aluminum oxide beneath the deposited carbon membrane can be due to oxygen migrating from the sapphire substrate through the defects in the nitride heterostructure (vacancies, threading dislocations) under the influence of the high temperature, or from the residual oxygen doping of the as-grown AlGaN layer (usually several $10^{18}/cm^3$ according to secondary ion mass spectrometry measurements).

In the following, high resolution scanning transmission electron microscopy (STEM) analyses combined with electron energy loss spectroscopy (EELS) are reported, to provide further insight on the structural and chemical properties of the near surface AlGaN region and of the deposited Gr membrane.

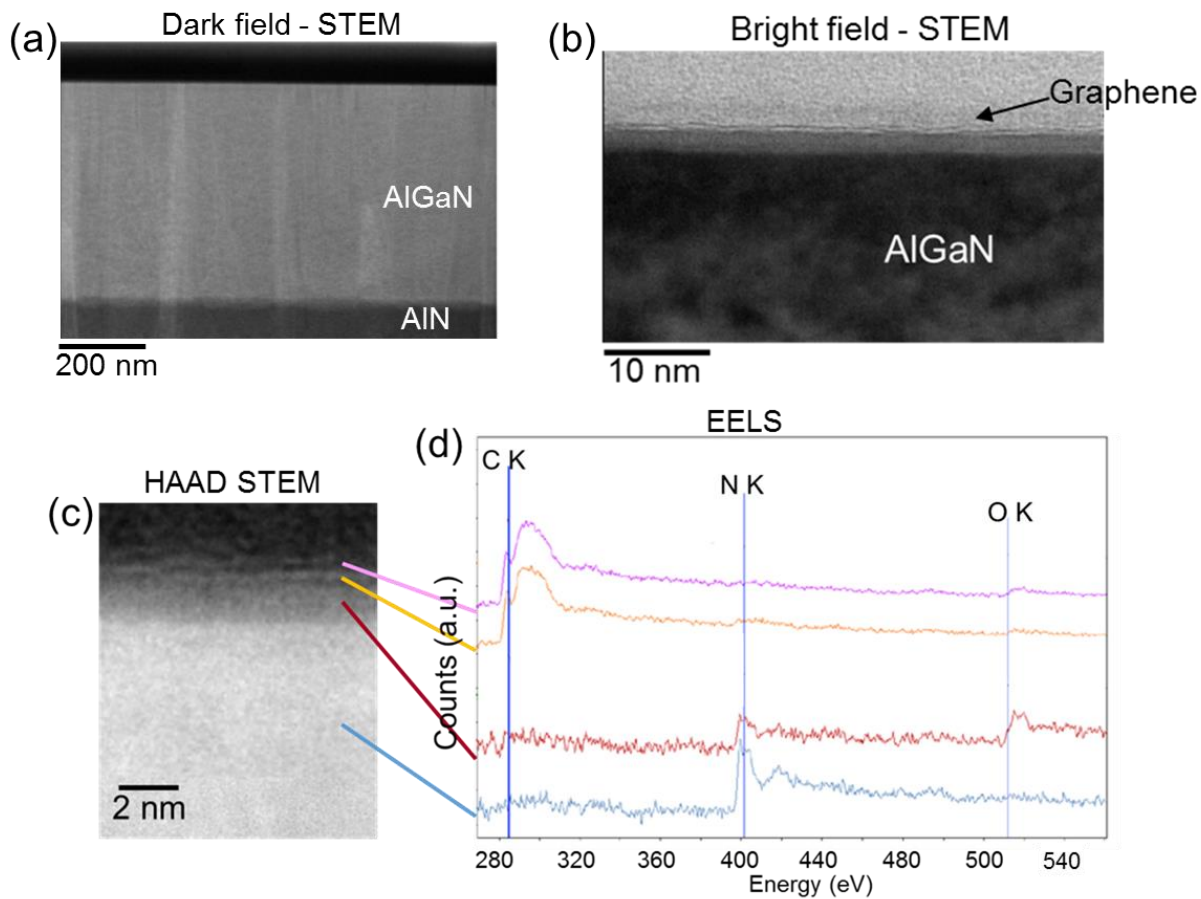

**Figure 6.** (a) Low magnification high angle annular dark field (HAADF) STEM image of the 500 nm thick AlGaN template, (b) Magnified bright field STEM image of the surface region, showing 2 layers of graphene on top of AlGaN. (c) HAADF-STEM taken at the interface between graphene and AlGaN (d) electron energy loss (EELS) spectra extracted at different positions, along the interface between AlGaN and the graphene bilayer, from the data set acquired within the field of view of the image showed in (c).

Fig.6(a) reports a low magnification dark field STEM image of the 500 nm thick AlGaN template, from which the presence of threading dislocations extending from the AlN layer to the sample surface can be deduced. A high-magnification bright field STEM image of the surface region is reported in Fig.6(b), where the presence of 2 layers of Gr on top of AlGaN can be clearly observed. Furthermore, it can be noticed that the topmost ~2 nm region of the AlGaN layer exhibits a different contrast with respect to the underlying part, which can be an indication of a different chemical composition of this near surface region. Such aspect is better elucidated by the high angle annular dark field (HAADF) STEM image in Fig.6(c) and by the EELS spectra reported in Fig.6(d). The contrast in the HAADF mode is sensitive to the atomic number (Z) of the species under investigations, with brighter signal corresponding to higher Z. The lower contrast of the 2 nm surface region with respect to the underlying AlGaN is consistent with a reduced concentration of Ga and partial oxidation of Al, as indicated by the XPS analyses in Fig.5. Finally, the EELS spectra in Fig.6(d) confirm oxygen incorporation in the near surface AlGaN region by the appearance of the O- K edge at ~510 eV, in addition to the N- K edge at ~400 eV. Furthermore, the graphitic nature of the bilayer on the surface is confirmed by the fine structure of the C K edge, which includes a sharp peak at ~285 eV associated to $\pi^*$ bonding and a broader peak at ~295 eV associated to $\sigma^*$ bonding –[45]. The intensity ratio of the $\pi^*$ and $\sigma^*$ peaks for our CVD Gr is lower than the one reported for high quality epitaxial Gr on SiC [45], consistently with the nanocrystalline nature of this material indicated by Raman spectroscopy.

In addition to the previous structural and compositional analysis, the electrical properties of the CVD grown Gr onto AlGaN have been investigated both on macroscopic scale, by four-point probe (FPP) measurements, and at nanoscale, using conductive atomic force microscopy (C-AFM).

Figure 8 reports a typical current-voltage characteristic measured in the FPP configuration (see schematic in the insert), i.e. by forcing the current ($I_{1-4}$) between two probes placed at two corners of a square Gr/AlGaN sample and measuring the potential difference ($V_{2-3}$) between the two probes on the opposite side. A low force is applied by the probes, in order to avoid punch through the ultra-thin Gr membrane. This curve exhibits a linear behavior, indicating an ohmic contact between the probes and Gr. A Gr sheet resistance $R_{sh} \approx 15.8$ k$\Omega$/sq was evaluated from the slope of this curve. Such resistance value is similar to the one reported for few layers nanocrystalline Gr deposited by direct CVD on other non-catalytic substrates, such as $SiO_2$ [46]. On the other hand, it is more than one order of magnitude higher than the typical values for unintentionally doped Gr grown by CVD on copper and transferred to insulating substrates [47].

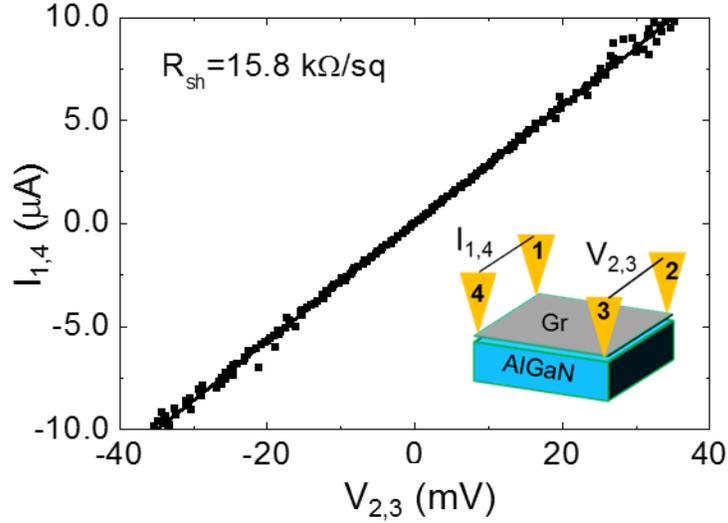

**Figure 7.** Typical four-point-probes (FPP) current-voltage characteristic measured on a square graphene/AlGaN sample (5 mm × 5 mm side). The insert illustrates the measurement configuration.

The nanoscale size of the crystalline domains and the large density of grain boundaries in our Gr obtained by non-catalytic CVD is certainly responsible for the higher sheet resistance as compared to the one of polycrystalline Gr membranes grown on copper, with reported domain sizes ranging from tens of micrometers to millimeters depending on the growth conditions [48]. However, in addition to its nanostructure, also mesoscopic scale inhomogeneities of the electrical properties (related to Gr thickness variations, wrinkles, and to the AlGaN roughness) can have an important impact on the overall sheet resistance of the Gr film. To this purpose, high resolution electrical characterization of Gr onto AlGaN was carried out by the C-AFM technique [49] using a conductive Pt tip as schematically depicted in Fig.8(a). In this configuration, the current flows laterally from the nanoscale tip/Gr contact inside the Gr layer, and it is finally collected by the large area contact deposited on Gr. Fig.8(b) and (c) show a typical morphological image and a current map of Gr surface on a 5 μm×5 μm scan area. The morphological image exhibits a roughness mainly associated to the AlGaN surface, overlapped with the wrinkles of the Gr membrane and some isolated carbon clusters. The current map shows lateral variations in the electrical properties of deposited Gr, which are partially correlated with the topographic features. The measured current values are mainly determined by the series combination, i.e. the sum, of the tip/Gr contact resistance and of the spreading resistance from the nano-contact to the Gr membrane [50]. Both contributions are related to the local carrier density in the Gr region underneath the tip. In particular, a locally reduced carrier density results in an increased tip/Gr contact resistance, as well as in a locally higher Gr resistivity, i.e. a higher spreading resistance [51]. Recent ab-initio simulation studies of the Gr interface with the AlN or aluminum oxide surface have demonstrated strong effects on Gr doping due to charge transfer phenomena [52]. The local changes in the Gr carrier density observed in the present case can be ascribed to local compositional variations in the near surface oxidized AlGaN region (as observed by STEM/EELS). Furthermore, a reduced Gr doping is expected in the wrinkle regions, due to the reduced charge transfer from the substrate in these corrugations of the Gr membrane [19].

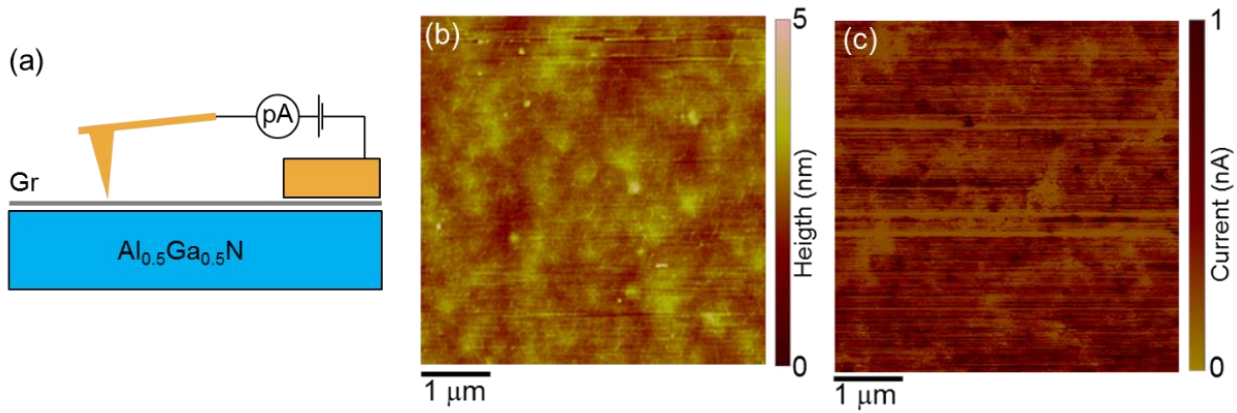

**Figure 8.** (a) Schematic of the experimental setup for C-AFM measurements on CVD grown graphene on AlGaN. Representative morphology (b) and current map (c) measured on the graphene surface by applying a tip bias of 100 mV.

## 5. Conclusions

In conclusion, we reported a detailed investigation of the morphological, chemical and electrical properties of few layers of Gr directly grown by CVD on Al-rich $Al_xGa_{1-x}N$ templates on sapphire. A preliminary assessment of the thermal stability of the AlGaN morphology in the temperature range from 1250 to 1450 °C (under $N_2$ ambient) revealed a similar behavior for $Al_xGa_{1-x}N$ templates with x=0.65 and 0.5 mole fractions. Uniform and conformal coverage of the AlGaN surface was achieved by $N_2/C_3H_8$ CVD at a temperature of 1350 °C, after optimization of the $C_3H_8$ flow rate. Interestingly, the Gr deposition was also found to prevent morphological degradation of the AlGaN morphology. However, Ga loss and partial oxidation of Al in the near-surface (~2 nm thick) AlGaN region was evidenced by XPS and cross-sectional STEM/EELS analyses. Raman spectra indicated that the deposited Gr membrane is nanocrystalline (with lateral domain size ~7 nm) and compressively strained, consistently with the results of STEM/EELS analyses demonstrating the presence of a bilayer of Gr with partial $sp^2/sp^3$ hybridization on top of AlGaN. Due to its nanocrystalline nature, the CVD grown Gr exhibits a sheet resistance of ~15.8 kΩ/sq, as determined by macroscopic FPP measurements. Furthermore, local variations of the Gr carrier density were revealed by nanoscale resolution C-AFM analyses, probably associated to changes in the charge transfer from the substrate due to local oxidation of AlGaN or to the presence of Gr wrinkles.

**Acknowledgments**

We acknowledge P. Prystawko, P. Kruszewski, and M. Leszczynski (TopGaN, Warsaw, Poland), I. Deretzis, A. La Magna, P. Fiorenza and R. Lo Nigro (CNR-IMM, Catania, Italy), F. M. Gelardi (University of Palermo) for useful discussions. S. Di Franco (CNR-IMM, Catania) is acknowledged for the expert technical support in samples preparation. This work has been funded, in part, by MIUR in the framework of the FlagERA-JTC 2015 project "GraNite" (MIUR Grant No. 0001411), by the FlagERA-JTC 2019 project "ETMOS", by the National Project PON EleGaNTe (ARS01_01007). CNR researchers thank the Italian Infrastructural project Beyond Nano Upgrade. CNRS researchers thank the French technology facility network RENATECH and the "Investissements d'Avenir" program ANR-11-LABX-0014.